\documentclass[table,prologue,smallextended]{svjour3}    
\smartqed 

\usepackage[utf8]{inputenc}
\usepackage{booktabs} 
\usepackage{float}
\usepackage{graphicx}
\usepackage[table,xcdraw]{xcolor}

\usepackage{breakcites}

\begin{document}

\title{The Impact of Looking Further Ahead: A Comparison of Two Data-driven Unsolicited Hint Types on Performance in an Intelligent Data-driven Logic Tutor}

\author{Christa Cody$^{1}$   \and
        Mehak Maniktala$^{1}$ \and
        Nicholas Lytle$^{1}$ \and
        Min Chi$^{1}$  \and 
        Tiffany Barnes$^{1}$
}


\institute{ Christa Cody \at
           \email{cncody@ncsu.edu} \\ \\
$^{1}$North Carolina State University, Computer Science Department, Raleigh, NC, USA \\
}
\date{Received and accepted date will be inserted later}

\maketitle

\raggedbottom

\begin{abstract}
    Research has shown assistance can provide many benefits to novices lacking the mental models needed for problem solving in a new domain. However, varying approaches to assistance, such as subgoals and next-step hints, have been implemented with mixed results. Next-Step hints are common in data-driven tutors due to their straightforward generation from historical student data, as well as research showing positive impacts on student learning. However, there is a lack of research exploring the possibility of extending data-driven methods to provide higher-level assistance. Therefore, we modified our data-driven Next-Step hint generator to provide Waypoints, hints that are a few steps ahead, representing problem-solving subgoals. We hypothesized that Waypoints would benefit students with high prior knowledge, and that Next-Step hints would most benefit students with lower prior knowledge. In this study, we investigated the influence of data-driven hint type, Waypoints versus Next-Step hints, on student learning in a logic proof tutoring system, Deep Thought, in a discrete mathematics course. We found that Next-Step hints were more beneficial for the majority of students in terms of time, efficiency, and accuracy on the posttest. However, higher totals of successfully used Waypoints were correlated with improvements in efficiency and time in the posttest. These results suggest that Waypoint hints could be beneficial, but more scaffolding may be needed to help students follow them.
\end{abstract}


\section{Introduction}
Intelligent tutoring systems (ITS) provide adaptive assistance to students and have significant positive effects on learning \cite{Murray1999, ma2014intelligent}. Multiple approaches to assistance have been explored, with some very specific assistance, like bottom-out hints \cite{vanlehn2006behavior}, designed to ensure that students ``do not flounder during problem solving'' \cite{merrill1992effective}, while other more abstract assistance, like a suggested subgoal \cite{catrambone1998subgoal}, is designed to allow more freedom and exploration within the domain. 

Providing assistance has been shown to reduce the cognitive load of learning by simplifying the task, leading to greater learning outcomes in less time \cite{kalyuga2011cognitive, sweller1988cognitive}. However, determining what level or type of help students need is a complex task that can affect learning outcomes \cite{aleven2000limitations, vanlehn2006behavior, wood1999help}. A major goal of providing assistance is to level the playing field of learning so that students at any incoming proficiency can master the same material in a similar amount of time. Research has shown that the level of hint and the learner's incoming experience can affect learning outcomes in ITSs \cite{arroyo2000macroadapting, kalyuga2011cognitive}. One example of this is the expertise reversal effect where methods that benefit novices, such as worked examples, become detrimental to students with higher expertise due to increasing cognitive load through redundant information \cite{sweller2008evolutionary}.

Furthermore, research has found evidence of aptitude-treatment interactions (ATI) within instructional strategies\cite{cronbach1977aptitudes, snow1991aptitude}, where certain students, particularly lower performin students, are more likely to be affected by changes in the learning environment. Similar to solving programming problems, solving logic proofs requires students to understand a system of domain principles or rules and to creatively apply them in sequence to achieve a goal. Consequently, support can be directed at any of these facets of problem solving, such as helping a student learn a rule or identify when applying a such a rule will move them towards a goal. Therefore, we hypothesized that different hint types could have different effects based on students' incoming proficiency. 

Deep Thought's default hints are Next-Step hints, where the next statement to be derived is given to the student and can be derived within one step. Providing the next step to derive allows students to focus their learning on discovering \textit{how} to reach their new short-term subgoal, rather than \textit{what} next subgoal to pursue. On the other hand, Next-Step hints may reduce student autonomy or practice in creating appropriate problem solving strategies. 

To evaluate the effect of a new hint type on student's performance, we created Waypoints, that can be thought of as \emph{intermediate subgoals}, by modifying our Next-Step hint generator to produce hints that mimic subgoals without the need for expert labelling. Our new method produces Waypoint hints that require students to perform 2-3 steps to derive them. Waypoints are intended to serve as near-term subgoals, that allow students more room for exploration and latitude in strategy construction.  

Another important aspect to assistance in tutoring systems is the ease of generation. Data-driven methods, where actions within the tutor are designed and developed using historical data, have been used to great effect to automate and individualize computer-aided instruction \cite{mostafavi2016evolution, stamper2013experimental, fossati2015data, barnes2008pilot}. Deep Thoughts's data-driven assistance matches current student work with similar historical successful, efficient examples to provide adaptive Next-Step hints using the Hint Factory, which is a method of generating Next-Step hints. The original Hint Factory opened a new field of data-driven hint generation that was first applied in tutors for logic \cite{stamper2008hint, barnes2008pilot}, and then for linked lists \cite{fossati2010generating}. More recently, the Hint Factory approach inspired new research in generating Next-Step hints for novice programming, based on generating assistance using pieces of previous student's solutions \cite{fossati2010generating, rivers2017data, price2017factors, barnes2008pilot}. 

However, there is a lack of research extending this Next-Step hint generation to provide additional forms of assistance. Therefore, we modified our Next-Step hint generator to produce Waypoint hints. Our modifications were inspired by the Approach Maps technique of graph-based mining to discover important subgoals in common student solutions \cite{eagle2014exploring}. This extension of Next-Hint generation to provide a higher level of hint may be used in other systems to easily generate a new hint type that could provide more adaptive assistance to address individual student needs.

Our goals for this study were to 1) perform a study to compare the impacts of Waypoints with Next-Step hints on performance, and 2) determine whether prior proficiency interacted with hint type to impact tutor posttest performance. We investigated the impact of the two types of hints, Next-Step and Waypoints, on student learning via unsolicited, tutor-initiated steps inserted into the student workspace, which we refer to as ``Assertions''. Assertions are designed to direct student attention to, and promote adoption of, unsolicited Next-Step and Waypoint hints.  
 
Based on the prior research mentioned above, we hypothesized that our Next-Step hints would be most beneficial for students with lower incoming proficiency and lead to better performance on the posttest. We also hypothesized that Waypoints would be more beneficial to students with higher incoming proficiency and lead to better performance on the posttest. In other words, we predicted an aptitude-treatment interaction (ATI) effect \cite{cronbach1977aptitudes, snow1991aptitude} where prior student proficiency would impact which students benefit most from a treatment. We predicted an ATI effect for both Waypoint and Next-Step hints, with higher proficiency students benefitting more from Waypoints and lower proficiency students benefitting more from Next-Step hints.

In this paper, we first discuss the context of the logic tutor, Deep Thought, and the method of generation for the different hint types. We then outline our experimental setup, designed to compare these two hint types in terms of their effects on student learning outcomes. Finally, we discuss the study results and how they relate to prior literature, and provide recommendations for future data-driven hint development and research.

\section{Related Work}
In this section, we discuss various approaches to assistance, such as subgoals, Next-Step hints, and worked examples, within intelligent tutoring systems (referred to here as ITSs or tutors). We also discuss cognitive theories surrounding assistance, including cognitive load and the ``zone of proximal development'', that have influenced our work.


Guided discovery, helping students discover new knowledge rather than providing direct instruction, is generally more beneficial than allowing students to learn unguided \cite{kirschner2006minimal, mayer2004should}. This finding agrees with the theory of the “zone of proximal development” (ZPD), the space between things a student can do independently and those they can only do with support \cite{vygotsky1978interaction}. Vygotsky hypothesized the most effective learning occurs when students are assigned tasks within their ZPD, meaning that tasks should neither be so simple that students can do them independently nor so difficult that they cannot make progress even with assistance. This dilemma of choosing an appropriate level of assistance shows that giving or withholding information is a delicate balance with trade-offs \cite{koedinger2007exploring}. 

The theory of cognitive load may explain the trade-offs of different approaches to assistance. Providing assistance can reduce the cognitive load needed for students to learn through methods such as simplifying the task \cite{kalyuga2011cognitive} or breaking the task down into easier-to-digest components, such as subgoals \cite{morrison2015subgoals}. However, the cognitive load of a learner is affected by both the elements of information in the task and their own ability \cite{sweller1988cognitive}. Intuitively, providing assistance that is too hard for a particular student to understand can negatively impact learning. However, providing assistance when it is not needed may also have a negative effect, such as the expertise reversal effect in which providing students information they already know increases their cognitive load \cite{sweller2008evolutionary}. On the other hand, it is a known problem that many students fail to request help when it is needed, and this has been termed hint avoidance \cite{aleven2000limitations}, discussed later in this section.

\subsection{Approaches to Assistance in Tutoring Systems}

Intelligent tutoring systems (ITSs) have significant positive effects on learning outcomes \cite{Murray1999}. Many forms of contextualized assistance have been explored in ITSs, such as hints, worked examples, and error feedback \cite{hume1996hinting, fossati2015data, ueno2017irt, vanlehn2006behavior}. The most minimal hint type is error-specific feedback, which provides a hint regarding an error the student has made \cite{vanlehn2006behavior}. Our tutor, as described below, includes basic error feedback when rules are not applied correctly. 

Many tutors use goal-directed hint sequences to provide several hints in a row, beginning with a more general hint then transition to more specific and directive hints \cite{hume1996hinting}. Our tutor has this capability, but it was disabled for this study to determine the impact of hint type and not the amount of detail each student might request. A standard goal-directed hint sequence within a tutoring system is Point, Teach, and Bottom-out \cite{hume1996hinting}. Pointing hints attempt to remind the student of relevant material. Teaching hints describe how to apply the relevant material. Bottom-out hints tell the student the next step and specifically how to implement it. The hints in Deep Thought would be considered pointing hints, because they point students in the direction they should be moving by giving them a hint statement to work towards.

\subsubsection{Higher-level Assistance}
One type of assistance higher-expertise learners benefit from is subgoals, a set of steps in the solution process that allows users to ``chunk'' information for ease of learning \cite{catrambone1998subgoal, morrison2015subgoals}. Sweller et al. \cite{sweller1982effects} found that using more abstract representations of goals in five maze-tracing experiments resulted in ``fewer errors and more rapid learning of the structure of the problem.'' The authors found that the more information solvers knew about the goal, the less they learned about problem structure. However, studies have found that these approaches have trade-offs depending on learner ability and problem difficulty or context \cite{morrison2015subgoals}. 

In regard to learner's abilities, research within ITSs has shown that high-ability learners can benefit from lower amounts, or less guidance, while low ability learners benefit from more concrete (specific and direct) guidance \cite{arroyo2000macroadapting, luckin1999ecolab}. These findings inspired us to explore how data-driven hint algorithms could be used to derive less direct guidance to benefit high-ability learners.
\subsubsection{Next-Step Hints}
The Hint Factory is a data-driven approach developed to generate Next-Step hints for students applying rules to solve open-ended problems in well-defined domains where there are multiple valid solutions \cite{stamper2008hint, stamper2013experimental}. New innovations in generating assistance from individual pieces of previous student's solutions have helped researchers extend the ideas of the Hint Factory to generate Next-Step hints for new domains including novice programming and linked list construction \cite{fossati2010generating, rivers2017data, price2017factors, barnes2008pilot}. The Next-Step hints derived by the Hint Factory and used in our tutor are pointing hints that suggest a statement a student could derive using a single domain rule application. Sweller et al. makes the case that providing more explicit instruction is better for novices who need to establish those individual learning blocks before they can create their own mental models \cite{sweller2008evolutionary,meerbaum2011habits}. 

However, research has shown that allowing students to make successful, unaided attempts at solving a problem can provide a higher learning benefit compared to explicit instruction showing them what to focus on \cite{koedinger2007exploring}. Hint Factory Next-Step hints have been shown to be successful in supporting student learning and problem-solving, with students having access to such hints in logic being 3 times more likely to complete the tutor than those without \cite{stamper2013experimental}. These results suggest that Next-Step hints are direct and explicit enough to support learning, but since level 1 hints do not provide the full information to achieve a next-step, students must do some unaided exploration to achieve the suggested hint statement. On the other hand, Aleven et. al notes a “one size fits all” strategy for guidance is not likely beneficial \cite{aleven2000limitations}. Hence, we are inspired to determine whether some even less direct data-driven hints may benefit high-ability learners.

\subsubsection{Aptitude-Treatment Effect}
Aptitude-treatment interactions have been widely studied in the educational domain. Prior research in instructional strategies \cite{cronbach1977aptitudes, snow1991aptitude} has shown the existence of aptitude-treatment interaction (ATI), where certain students are more sensitive to variations in the learning environment and may be affected differently by the treatment compared to less sensitive students who perform regardless of the treatment. Educational researchers have discovered ATI effects based on prior experience level, prior working memory, and incoming self-regulated learning ability \cite{kalyuga2001problem, lehmann2016working, fuchs2019using, yeh2015aptitude}. For example. Lehmann et al. explored the effect of working memory on learning outcomes in fluency/disfluency groups, where instructional materials had different levels of text legibility \cite{lehmann2016working}. Based on these findings, we believe that there could be an aptitude-treatment effect associated with hint type. We believe that students with lower incoming proficiency may be more sensitive to hint type. 

\subsubsection{Help Avoidance and Unsolicited Hints}
Despite this considerable research on assistance, there is pervasive problem within ITSs called help avoidance, where students do not leverage the intelligence within the system for help \cite{aleven2004toward}. There are many reasons for help avoidance, one of which is that certain students may lack specific meta-cognitive skills like knowing when to ask for help \cite{aleven2000limitations}. As a result, some ITSs employ unsolicited hints (i.e. providing hints when needed without request) to prevent help avoidance \cite{murray2004looking}, and we adopt this unsolicited strategy here.

Zhou et al. found that students were more likely to make effective pedagogical decisions at the problem-level rather than the step-level, meaning that students were less able to make effective decisions when deciding if they need a hint on a particular problem-solving step \cite{zhou2016impact}. In another study, researchers found that a large number of students using Andes, the physics tutor, would guess instead requesting hints \cite{ranganathan2014students}. Furthermore, higher learning gains have been observed for low performing students when unsolicited hints were provided\cite{arroyo2001analyzing}. While one study found that students learned more reliably with hints on-demand than unsolicited hints\cite{razzaq2010hints}, other studies have shown that providing hints at the appropriate time can augment students’ learning experience \cite{bunt2004scaffolding,puustinen1998help}, improve their performance \cite{bartholome2006matters}, and avoid the negative effects of frustration while saving students time by preventing unproductive struggle \cite{murray2006comparison}. 

Within our tutor, even though students often have difficulty and hints are readily available via the hint button, most students do not request assistance. In Fall 2017, students using our tutor requested a median of zero hints per problem. In this study, to enable us to compare the impact of hint type, we periodically (frequency defined in Section \ref{sect3:assist}) provided unsolicited hints to students based on the condition they were assigned. In prior work, we compared our unsolicited hints to the normal conditions in Deep Thought, on-demand hints only, and found that the unsolicited hints had no impact on the performance metrics in the training and no negative impacts on any performance metrics on the posttest \cite{codydoes}. Furthermore, this work found that providing unsolicited hints reduced steps that students needed help, but didn't receive it as detected by our Help-Need model \cite{mehak, maniktala2020extending}. Therefore, we do not believe that our unsolicited hints are disruptive, but we note that providing unsolicited hints has potential for disrupting students' learning. In the next section, Deep Thought and its interface are discussed in detail and the hints' generation, usage, and frequency are expanded on.

\section{The Deep Thought Logic Proof Tutor}

\begin{figure}[ht]
\centering
\includegraphics[width=\textwidth]{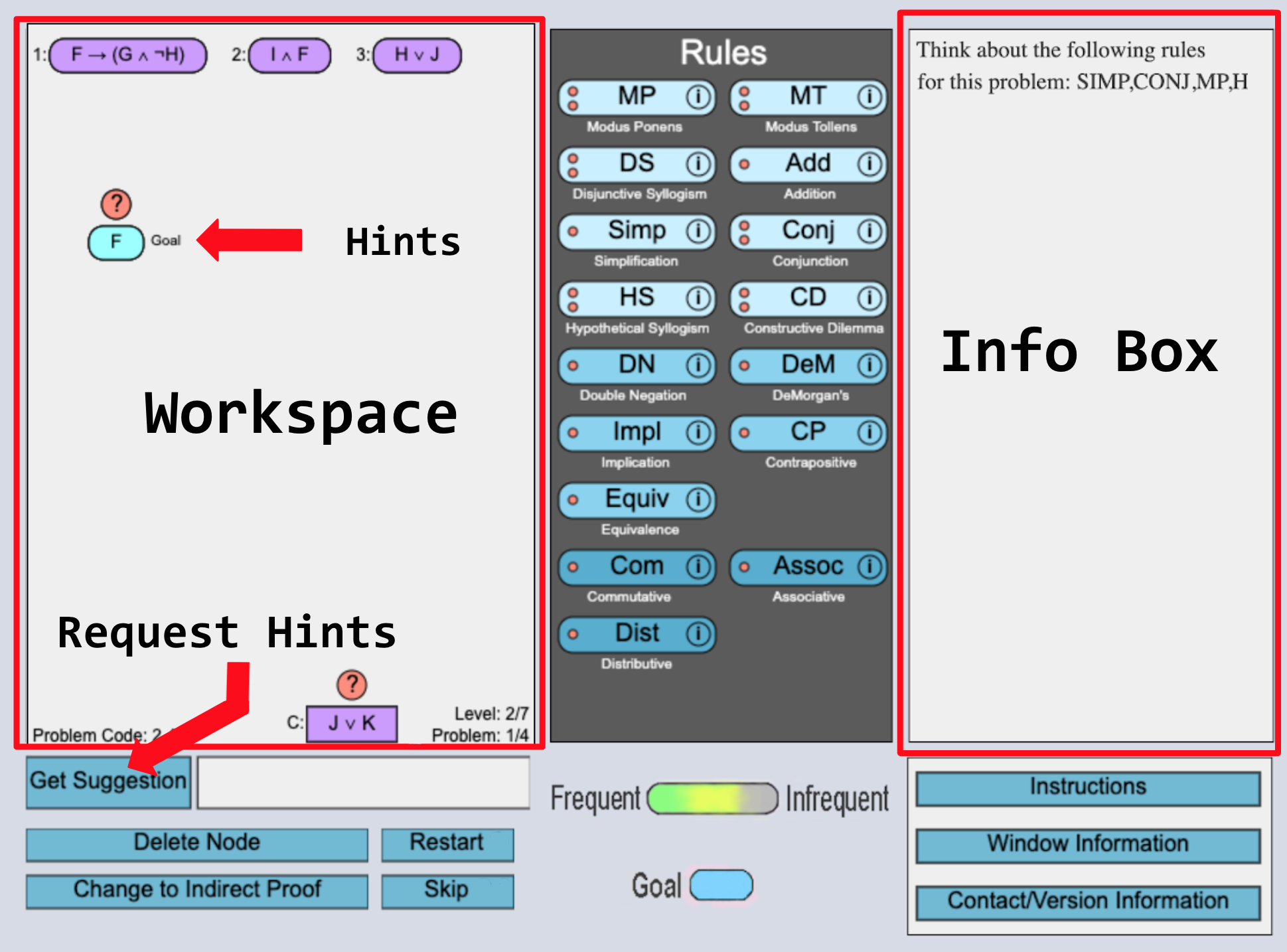}
\caption{On the left of the screen is the Deep Thought workspace. Below the workspace are are the hint button and hint message box, the rules are in the middle, and to the right is the Dialogue Box where messages related to unsolicited hints as well as problem information are given.}
\label{fig:Screen}
\end{figure}

The Deep Thought tutor (see Figure \ref{fig:Screen}, described further below) is used in the context of a discrete mathematics course where students first spend 2 weeks learning about truth tables, and proving each logic rule is true in class and in online multiple-choice homework assignments. Then, students learn about formal proofs, where students iteratively apply logic rules to a set of given statements to derive a specified conclusion. 

A formal proof works much like any multi-step procedural problem where domain principles are applied to given and previously-derived facts to derive and justify new statements. For example, in physics, students may be given values for mass and acceleration and be asked to determine force. They would then apply the domain principle of $F = m * a$ along with the given values of $m$ and $a$ to derive a new statement about the value of $F$. In logic, each derived statement must have a justification which consists of the domain principle and the relevant prior statements it was applied to. This corresponds to the information used to derive $F$ in the previous physics example. In a formal proof, students are \textbf{given} a few statements (the number may vary) that are known to be true - often referenced as ''givens'' - and a conclusion that is to be derived. Then, students must \textbf{apply} logical rules to the givens to \textbf{derive} new statements. The student repeats this process of identifying rules to apply on certain statements until they derive the conclusion. An example of this process in Deep Thought is covered in this section along with a description of the interface.

Within the discrete math course, students next complete partially-worked examples in a fill-in-the-blank type interface where they are given formal logic proofs with one missing part on each step - either the derived statement, or part of the justification that consists of the rule used to derive it and the statements the rule was applied to. Many example logic proofs are worked in class, with students asked to actively solve logic proofs in small groups, and students are provided with several full worked examples in handouts. After this class work and homework, students are assumed to have reasonable familiarity with logic rule application, but need practice in determining which rules to apply in service to a problem-solving goal. Students are then assigned to complete formal logic proofs using our propositional logic tutor called Deep Thought \cite{mostafavi2017evolution}. 

The intention of the Deep Thought tutor is to provide students with practice on solving logic proofs with a focus on problem-solving efficiency in both time and the number of steps in their solutions, i.e. shorter proofs in less time, and ideally with few mistakes in justifying or deriving new statements. To do so, the tutor must provide basic functionalities including (1) correctness feedback on each step (on both justification and derived statements), and (2) automated detection of proof completion. Like a compiler, Deep Thought provides these functions that identify errors and clearly shows when a problem is \textit{complete} but do little to help students with the overall goal of reaching a problem solution through deriving and justifying a series of well-chosen statements. To bridge this gap, the Hint Factory was created to provide data-driven assistance that could point students to appropriate subgoal statements to derive \cite{barnes2008pilot, stamper2008hint, stamper2013experimental}.

Deep Thought allows students to solve logic proofs graphically as shown in Figure \ref{fig:Screen}. On the left of Figure \ref{fig:Screen}, the workspace is labelled. The workspace is where the students can select statements (purple, oval-shaped nodes) and apply rules by selecting rules (blue, oval-shaped nodes) from the middle of the screen under the ``Rules" section to derive new statements. In Figure \ref{fig:Screen}, there are 4 givens (at the top of the workspace in purple, oval-shaped nodes) and the conclusion (at the bottom of the workspace in a purple, square-shaped node). Each statement is labelled to show the order in which students derived them with the exception of the givens and conclusion which are labelled for ease of reference. There is no particular ordering to the givens. Also, there is an example of our hints on the screen in the blue, oval-shaped nodes labelled ``Goal."

To derive a new statement, a student must select statement(s) by clicking them followed by selecting a rule to apply. In response to the student selecting a rule to apply, the tutor has one of 3 responses: 1) if the student is using an applicable rule, i.e. a rule that logically can be applied to the statements, and the new derived statement is the only potential derivation, then the statement is automatically added to the screen, 2) if the student is using an applicable rule and there are multiple potential derivations, e.g. using ``Simplication" on the statement ``$I\land F"$ where either I or F could be the new derived statement, then the student is prompted to enter in the statement they want to derive, and 3) if the student has incorrectly selected a rule that doesn't apply to the selected statements, e.g. the rule requires only one statement to be selected, such as ``Simplification" or ``Implication", and the student has selected two statements, then the tutor provides a pop-up and a description of the error. Note that in response 2, if the student incorrectly types in the statement to be derived in the prompt, the tutor will pop-up and error and the student will have to select the rule again to derive the statement. 

When a new statement is added by the student, the statement becomes oval-shaped node, similar to the givens shape, but the color depends on the frequency and necessity of the node based on historical data. To help students avoid deriving unnecessary statements/nodes in the training phase, the tutor colors nodes based on their necessity and frequency in our historical dataset of correct solutions by past students. Nodes that were never necessary to derive the conclusion are colored gray, while frequently-necessary nodes are colored green, and infrequently-necessary nodes are colored yellow. 

As the student is deriving new statements, the nodes are added to the proof with arrows pointing to them to show which statements were used and the rule applied to derive the new statement. On the right of the screen, the Info Box contains information communicated by the tutor about the current problem the student is solving, i.e. what rules may be useful to solving the proof, information about hints on the screen, and information about certain buttons a student may try to use. The bottom left of the screen under the workspace contains buttons that a student may use during the training portion of the tutor (skipping a problem and requesting suggestions, i.e. hints, are not available during the testing portions of the tutor). To the bottom right of the screen, buttons are available that show the student general information about the tutor as well as instructions that provide information about solving proofs and the options available for the students.

As stated above, Deep Thought is intended to teach students to solve proofs more efficiently, in terms of time and steps taken to reach the conclusion. The tutor presents proof problems as an initial set of given statements with a conclusion to derive from them using logic rules. Each statement, given or derived, is represented by a \textit{node}, with the conclusion represented with a node with a question mark `?' above it, indicating that it has not yet been justified (shown to be true using logic rules applied iteratively to the givens). 

\begin{figure}[!h]
\centering
\includegraphics[width=.5\textwidth]{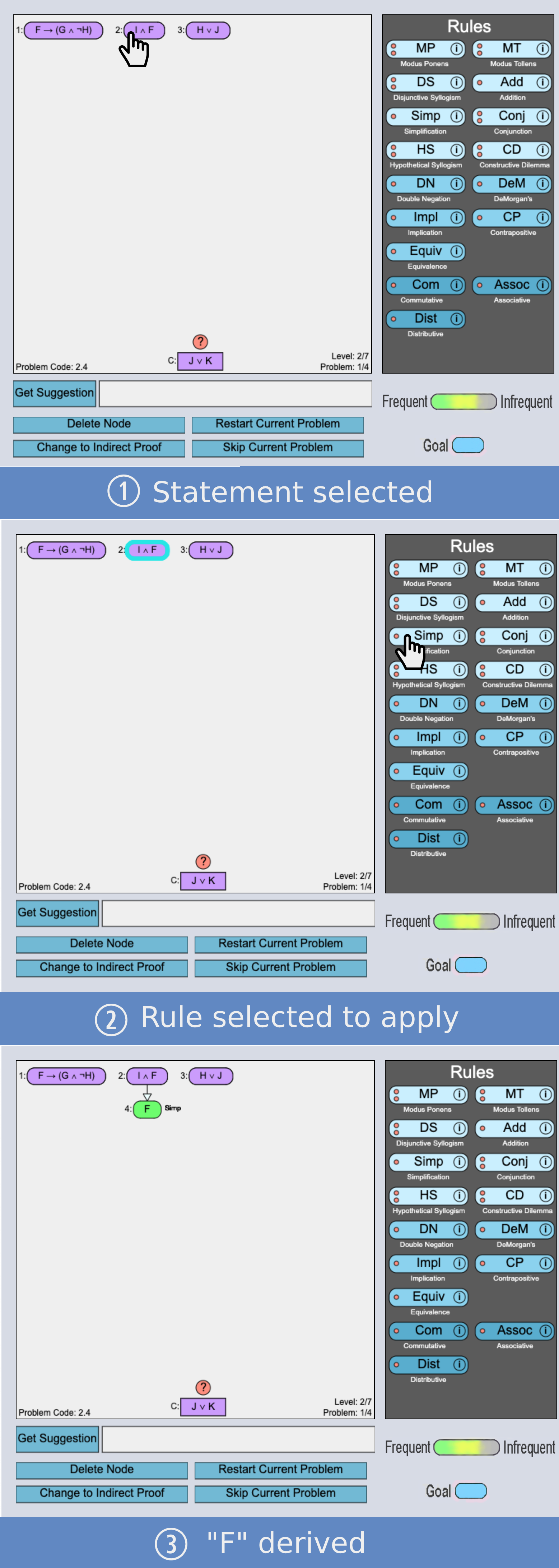}
\caption{Deriving a new justified node. (1) Selecting the node ``$I\land F"$ to use (2) Selecting the rule ``Simplification" to apply (3) The screen after the rule was clicked showing ``F" as a justified node }
\label{fig:Derivation}
\end{figure}

Each problem-solving step consists of two parts: the justification and the derived statement. The justification is the set of 1-2 existing nodes and the rule applied to them, and the derived statement is the result. Students complete the justification by clicking to select 1-2 nodes, and clicking on a rule to apply. Students then type in the derived statement that results from applying the rule to the selected statement nodes. For example, Figure \ref{fig:Derivation} shows a formal proof beginning with 3 givens (at the top of the workspace in purple, oval-shaped nodes) and the conclusion (at the bottom of the workspace in a purple, square-shaped node). The student (1) selects the statement ``$I\land F"$ and (2) applies the ``Simp" rule, i.e. Simplication, to (3) derive a new statement ``$F$". To solve the proof, the student would continue identifying combinations of statements and rules to apply until they derive the conclusion statement ``$J\land K$". 

Throughout the tutor, including the pre- and post-test problems, Deep Thought provides immediate error feedback for mistakes - either in justifications or derived statements. If a student clicks on the wrong rule, or their derived statement does not follow from the selected nodes and rule, Deep Thought shows a popup message and records the error. For example, if a student selects two nodes and then clicks on the $Simp$ rule, the error prompt reads ``Rule requires one premise," then fades away. If the student enters a derived statement that is true, and the justification (consisting of the selected nodes and rule to derive it) is correct, then a new node with the derived statement appears in the workspace. 

To complete a problem, the student must iteratively derive and justify new statements, until the conclusion statement is derived and justified. When students have completed a problem, the conclusion's question mark is removed, and it is visually connected to the givens through a series of derived nodes and arrows indicating their justifications. Since the system automatically checks each step and detects completion in all phases of the tutor, student solutions cannot be incorrect, but some may be more expert than others. Students are considered to have learned the topic when they perform well on the posttest (described below), especially with regard to problem solutions with fewer steps and fewer mistakes in less time.

Deep Thought includes four phases: introduction, pretest, training, and posttest. The introduction consists of three problems including two worked examples, where students click through the addition of successive nodes until a conclusion is derived, and a third problem students solve alone to learn the interface. Then, students take the \textbf{pretest} consisting of a solving single problem with no hints available. The pretest is used to measure students' incoming proficiency and assign them to conditions via stratified sampling. Next, students solve 18 problems in the \textbf{training} section. For each training problem, the dialogue box provides information on what rules to focus on while solving a problem, such as ``Think about the following rules for this problem: MP, Simp, Add." Students also receive contextual, data-driven hints during training, including both unsolicited hints generated by the system and on-demand hints upon student request, all generated using the same Hint Factory-type approach described below. After completing training, students take a more difficult, non-isomorphic \textbf{posttest}, where they must solve four problems without any help or assistance. Since the posttest is not isomorphic to the pre-test, we do not expect the post-test performance to be directly comparable to the pretest performance. Rather, we use the pretest to balance incoming proficiency across groups via stratified sampling, and focus on comparing post-test performance between groups. 

Expert solutions for all tutor problems range from 5-8 steps, and student solutions typically contain 5-20 steps. Longer student solutions may simply be inefficient, taking more steps than needed, or they may contain unnecessary nodes that do not lie on a direct path from the givens to the conclusion (see footnote \footnote{Unnecessary nodes in a complete solution are easy to detect because removing them does not disconnect the conclusion from the givens, but they are difficult to detect during problem solving.}). As first mentioned in Section 3, to help students avoid deriving unnecessary statements/nodes in the training phase, the tutor colors nodes based on their necessity and frequency in our historical dataset of correct solutions by past students. Nodes that were never necessary to derive the conclusion are colored gray, while frequently-necessary nodes are colored green, and infrequently-necessary nodes are colored yellow.

\subsection{Assistance}
\label{sect3:assist}
In \textit{training problems}, students may receive unsolicited hints, depending on their assigned study condition, as well as request on-demand hints. On-demand hints and unsolcited hint provide the same content. All hints provide a target statement to derive, appearing as a node with a `?' in the workspace. In previous work, we showed this method of providing unsolicited hints, Assertions, resulted in better performance than text-based messages as a method of unsolicited hint delivery \cite{mehak}. For the remainder of the paper, we refer to both solicited and unsolicited hints as hints.

Deep Thought includes several measures intended to prevent gaming the system, where students attempt to use system features to avoid work, or help abuse, where students request hints when they do not need them \cite{baker2004detecting}. First, whenever a hint is already in the workspace, students may not receive another hint, whether it was solicited or provided automatically by the tutor. Second, no further details are provided for any hint, meaning there is no such thing as a bottom-out hint in this study. In past Hint Factory implementations, we have provided students 4 levels of hints that (1) suggested the next step, (2) the specific rule, (3) the prior statements needed, and finally (4) a bottom-out hint with all this information. In this study, we use only level 1 pointing hints, and disabled hint levels 2-4. 


 The tutor generates hints using historical student data from four semesters, each semester with approximately 250-300 students using the tutor. Both hint algorithms produce assistance based on the most \emph{frequent} and \emph{efficient} paths available in the student's current proof.
 
 We use the Hint Factory \cite{stamper2008hint} approach to generate hints. The Hint Factory \cite{stamper2008hint, stamper2013experimental, barnes2011using} is a data-driven method to generate hints by transforming historical student problem-solving attempts into a Markov Decision Process, using observed frequencies as transition probabilities, and estimating the expected value of each previously-observed problem state based on assigning rewards to complete solutions, small negative rewards (i.e. costs) to steps to positively reward more efficient solutions, and large negative rewards to errors to de-emphasize solutions that cause many students to make mistakes. Individual student problem-solving attempts are represented by a series of states, or snapshots of the work done so far, where transitions occur between states when students add or delete problem nodes, or make an error. The Hint Factory is described in detail in Barnes and Stamper's chapter in the 2011 Handbook on Educational Data Mining \cite{barnes2011using}. All student solutions are combined into an interaction network \cite{eagle2012interaction} that reflects all previously-observed solutions to one specific problem. When a hint is requested by the student or tutor, the Hint Factory is used to select a target problem-solving state with the highest expected value. Note that this process can be done offline, and a simple table can be used to store problem-solving states and their corresponding hint content for real-time hint provision. Then, the latest statement derived in that state is used as the pointing hint to help students know \textit{what} to try to derive next. 
 
 In this study, we do not provide further information on \textit{how} to derive or justify the suggested statement, i.e. the statements that a student needs to select and the rules the student may need to apply are not provided to the student, meaning that all hints in this paper can be considered as partially-worked example steps.
 
\begin{figure*}[!ht]
\centering
\includegraphics[width=.5\textwidth]{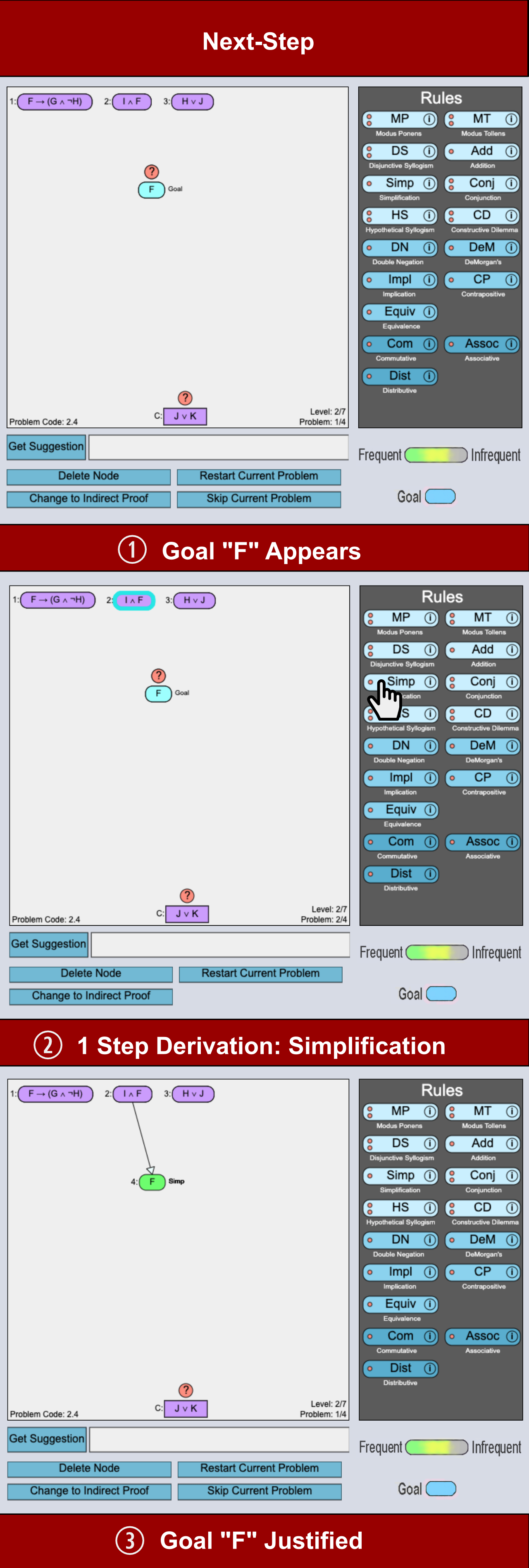}
\caption{Next-Step hint. (1) A Next-Step hint appears, $F$. (2) The student has selected $I\land F$ and is applying the Simplification rule. (3) $F$ has been justified.}
\label{fig:HintTypes_NS}
\end{figure*}

\begin{figure*}[!ht]
\centering
\includegraphics[width=.47\textwidth]{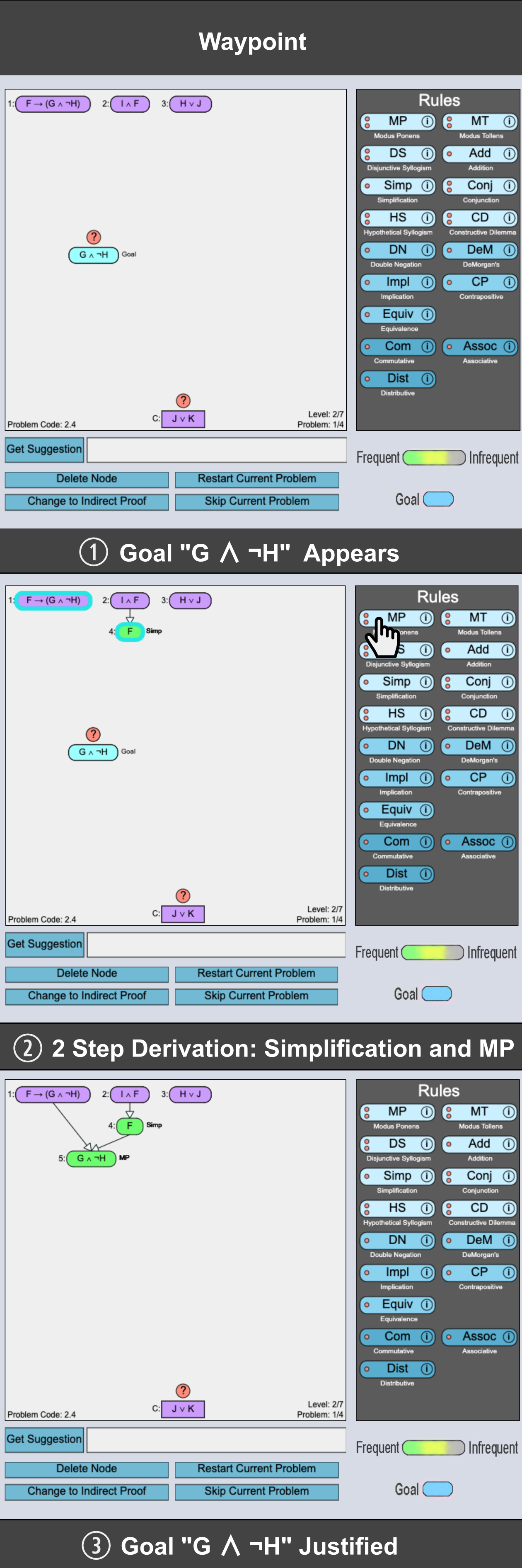}
\caption{Waypoint hint. (1) A Waypoint hint appears, $G \land \lnot H$. (2) The first derivation using Simplification has already been completed. The student has selected $F \rightarrow (G \land \lnot H)$ and $F$ and is applying \textbf{MP} (Modus Ponens). (3) $G \land \lnot H$ has been justified.}
\label{fig:HintTypes_WP}
\end{figure*}

 In this study, two hint types are used: \textbf{Next-Step} and \textbf{Waypoint} hints. Figure \ref{fig:HintTypes_NS} and Figure \ref{fig:HintTypes_WP} shows the two forms of data-driven hints: Next-Step (NS) and Waypoints (WP), respectively, and how the students would approach deriving the suggested hint statement for each type. Descriptions of how each hint type is generated as well as how to derive each hint are expanded in the following paragraphs.

\textbf{Next-Step hints} are generated using the Hint Factory method as described above, with the target state selected to be the one with the highest expected value that occurs \textbf{within one rule application from the student's current state}. Simply, Next-Step hints suggest the best proposition that can be derived in one step from the student’s current proof. This corresponds to the next-step hints derived in all of our prior work using Hint Factory \cite{barnes2008pilot, barnes2011using, stamper2013experimental, stamper2008hint, eagle2014exploring, eagle2012interaction, price2017factors, price2017hint}. 

Since Next-Step hints are partially-worked, they allow students to focus on \textit{how} to justify them, and reflect on \textit{why} they were suggested. This removes a considerable load; without a hint, students must also search among many options for the best \textit{what} to derive next. For example, Figure \ref{fig:HintTypes_NS} demonstrates the ideal derivation of a Next-Step hint. In Figure \ref{fig:HintTypes_NS}, Deep Thought has 3 givens at the top of the workspace and one hint statement labelled ``Goal", $F$, on the screen. To derive the hint, the student (1) selects the $I\land F$ statement by clicking it. After selecting the statement, which is now shown highlighted in blue, the student clicks the rule labelled ``Simp" to apply Simplification to the statement. A pop-up will appear for the student to type in what they are attempting to derive, in which case they enter $F$. After entering $F$ into the prompt, (3) the statement is shown incorporated into the student's solution in the same fashion regular derivations happen as in Figure \ref{fig:Derivation} with arrows, coloring, and labelling. In this case, the justified hint appears on the screen as a green, oval-shaped node with an arrow pointing to it from the $I\land F$ statement with the labels ``4:" to indicate this statement is the fourth statement justified (givens are automatically numbered) and the label ``Simp" to indicate the statement was derived using the Simplification rule.

\textbf{Waypoint hints} are generated with the same method as Next-Step hints; however, instead of selecting a hint 1 step away from the student's current state, hints that are 2-3 steps away from the student's current state are selected. A primary motivation for this study was to determine a simple way to extend the Hint Factory to provide less direct data-driven hints, i.e. compared to Next-Step hints, without the need for expert authoring. In our prior work, we derived a new method called data-driven Approach Maps, that applies hierarchical graph mining to interaction networks to discover problem-solving states that represent critical junctures in problem-solving attempts, which we call subgoals \cite{eagle2014exploring}. These subgoals occurred every 2-3 steps/states in our short logic proofs (which are typically 5-12 steps long). These subgoals inspired our Waypoints, but we wanted to be able to generate these hints with an easier method that is more extensible to other researchers who may already be using Hint Factory or methods based on Hint Factory.

To generate Waypoints without the need to apply data-driven Approach Maps, we modified the Hint Factory to select a target statement that was 2-3 steps away from the current state. Among states that were 2 or 3 steps away, we selected the state with a higher frequency within prior correct solutions. This resulted primarily in states that need only two rule applications to derive, since the diversity of student solutions means that frequency typically decreases in interaction networks the further states are from the start. By expert review of a random sample of Waypoints, we verified that this simple algorithm results in similar hints to those generated using data-driven Approach Maps \cite{eagle2014exploring}.
 
Waypoints are intended to serve as \textit{subgoals}, giving students more room to explore the solution space and develop their own problem-solving strategies. Since Waypoints cannot be achieved with a single rule application, they require students to make their own problem-solving plan to derive them, considering the existing problem statements and how rules might be applied to them to derive and justify the suggested Waypoint statement. For example, Figure \ref{fig:HintTypes_WP} demonstrates the ideal derivation of a Waypoint hint. In Figure \ref{fig:HintTypes_WP}, (1) Deep Thought has 3 givens at the top of the workspace and one hint statement labelled ``Goal", $G \land \lnot H$, on the screen. To derive the hint, the student first selects the $I\land F$ statement by clicking it, then the student clicks the rule labelled ``Simp" to apply Simplification to the statement. A pop-up will appear for the student to type in what they are attempting to derive, in which case they enter $F$. Note, this step is not shown in the figure, although it is the same process as described in Figure \ref{fig:HintTypes_NS}. After entering $F$ into the prompt, the $F$ statement is shown incorporated into the student's solution. Next, the student must make a second derivation to derive the hint. The student (2) selects the $F \rightarrow (G \land \lnot H)$ statement and the $F$ statement by clicking each one individually -- this highlights both nodes -- then the student clicks the \textbf{MP} rule to apply Modus Ponens to the statements. As a result, the statement is automatically derived, due to the derived statement being the only option, and the justified hint appears on the screen as a green, oval-shaped node with arrows pointing to it from the $F \rightarrow (G \land \lnot H)$ statement and $F$ statement with the labels ``5:" to indicate this statement is the fifth statement justified (givens are automatically numbered) and the label ``MP" to indicate the statement was derived using the Modus Ponens rule. 

For both Next-Step and Waypoint hints, the process of deriving the hint is the same: students must select statement(s) and apply rules to derive new statements, which we also refer to as ``steps". The only difference is how many times a student must repeat this process to derive the hint statement (ideally once for Next-Step and twice for Waypoint hints - with the exception of some Waypoints that may take three steps to derive). With that in mind, students may also derive new statements that do not contribute to deriving the hint statement; however, when we refer to how many steps it takes for a student to derive a hint, we are speaking in \textit{ideal} terms. 

We consider a continuum of goals for students, where Next-Step hints ideally take one step to derive, Waypoints take 2-3, and the problem conclusion takes about 5 expert steps. With longer problems or more complex problem domains like programming, we would recommend using a more complex algorithm to select Waypoints if they were shown to be effective. In logic proofs, the shortest proof is considered to be the best, so simple metrics on interaction networks can quickly discover optimal solutions and those that many students can discover.

As stated above, Deep Thought only provides pointing hints to suggest statements that can be derived; neither Next-Step nor Waypoint hints tell students which rules to use to derive them; rather, they help students solve problems by suggesting a subgoal that helps them break down multi-step problems. To use a hint in their proof, the suggested hint statement must be \textit{justified} by applying a rule to previously-justified or given statement(s). Statements that are not justified appear in the tutor interface with a ``?" above them to indicate that they need to be derived.

We implemented unsolicited hints so they appear randomly and with enough uniformity and frequency that even students with short proofs would receive hints. One limitation of this method of providing hints is that hints were not necessarily provided when they were most needed, which may affect learning outcomes. However, since students in tutor rarely request hints, it was necessary to provide the hints automatically and frequently to enable us to evaluate our hypotheses. For the Next-Step group, we capped the number of unsolicited hints at $1/3$ of the problem length and checking every 2-3 steps to see if a hint was still extant in the workspace (e.g. it was not yet justified). If a hint still remained, the algorithm did not provide a new hint, but if there were no hints on the screen, a new one was provided. Since Waypoints take more steps to derive, they remained unjustified for longer, and thus resulted in fewer Waypoint hints by design. Note that students can delete problem nodes at any time (excluding the givens and conclusion), and this includes hint nodes, even if they are not yet justified.


\section{Methods}
The Deep Thought tutor was used as a homework assignment for an undergraduate `discrete mathematics for computer scientists' course in the Fall 2018 semester at a large research university. We analyzed 143 students' data from two test conditions to investigate the impact of hint type on student performance and behavior. Both conditions were identical except for hint type, Next-Step or Waypoint. We used stratified sampling based on pretest performance, then randomly assigning to Next-Step hints (NS, $n$ = 71) or Waypoints (WP, $n$ = 72), ensuring both conditions were balanced in incoming knowledge. Before analysis, students who dropped the tutor before completion and students with technical errors in their data were removed (NS $n$ = 15, WP $n$ = 14) leaving 56 students in the NS condition and 58 students in the WP condition for a total of 114 students. 

\subsection{Hypotheses}
The goals of this study were to 1) evaluate the effectiveness of a new hint type, Waypoint hints, 2) compare the impacts of Waypoints and Next-step hints on performance, and 3) determine if proficiency had an effect on which hint type was more beneficial. Based on prior literature, we developed the following hypotheses: 
\begin{itemize}
  \item $H_{1}$: Next-Step hints will improve performance for students with lower incoming proficiency.
  \item $H_{2}$: Waypoint hints will improve performance for students with higher incoming proficiency.
  \item $H_{3}$ Waypoint hints will be more difficult to derive, resulting in a lower justification rate and performance during training compared to Next-Steps. 
\end{itemize}
These hypotheses were based on the basic assumption that Waypoint hints are more difficult to justify and adopt, since Waypoints require students to derive more steps to justify them. On the other hand, this challenge may be precisely what high-proficiency students need for improved learning. To evaluate these hypotheses, we focused on the performance metrics discussed below.

\subsection{Performance Evaluation Metrics}
In this section we describe the metrics used to evaluate student performance. Recall that the tutor begins with an introduction with two worked examples and one practice problem followed by the pretest. We used each student’s pretest \textbf{score} to measure incoming knowledge/proficiency. Equation \ref{eqn:score} shows how the score is calculated. Each metric is normalized, then the time and step metrics are subtracted from 1 to be comparable to accuracy, i.e. so that for time, steps, and accuracy a number closer to 1 indicates the student is performing well. A student's score is a combination of percentiles for the pretest \textit{time}, number of \textit{steps}, and \textit{accuracy} on a single problem, ranking students based on how fast, efficient, and accurate they are compared to their peers. We chose these features because they each represent a different aspect of a student's problem solving experience. 

Recall that the tutor was designed to improve time and steps to solve problems, and assumes a basic level of fluency or accuracy on rule applications. Therefore, we have no goals or expectations of improving accuracy with this tutor. However, the score includes all three metrics to ensure that our interventions do not decrease accuracy while attempting to improve time and steps. For example, a student may take a short amount of time on a problem, but make many mistakes resulting in a lower accuracy. We use a median split on the combined pretest score to assign students into High and Low proficiency groups for some analyses. 

\begin{equation}
\label{eqn:score}
Score = (1-TotalTime)*.5 + (1-TotalSteps*.3) + Accuracy*.2
\end{equation}

We investigated pre- to posttest changes as well as performance impacts on time spent solving a problem, total attempted steps, and accuracy. \textbf{Total time} is counted from the moment a problem begins until it is solved by deriving and justifying the conclusion. \textbf{Total steps} in a problem include any attempt at deriving a new node, which includes correct and incorrect steps. \textbf{Accuracy} is the percentage of correct out of the total steps, which is expected to start relatively high due to prior exposure in the class, and increase as students practice. Note that the tutor is not designed or assumed to promote large improvements in accuracy, since no penalties are assigned for incorrect rule applications and the tutor simply alerts students upon wrong rule applications and students may try again, even within the pre- and post-tests. Further, problems require new rules and become more difficult as the students progress. As we seek primarily to promote more efficient problem solving, we focus more on steps and time per problem while maintaining reasonable accuracy. This is because it is more difficult for students to learn to determine which steps to derive to achieve shorter, more efficient proofs, compared to learning how to apply the rules, which can be done by memorization and simple practice. Deep Thought is built primarily to allow students to practice with the strategy of problem solving, rather than fluency with rules, most of which are assumed to be learned before the tutor.

One important thing to note is that Deep Thought does not include eye-tracking, and the unsolicited hints are provided regardless of whether a student needs them or not, so we cannot determine precisely whether students followed a hint or incidentally derived the hint statement. Therefore, we have defined metrics to quantify when students \textit{justified} a hint by selecting the statements and rule needed to derive it, as well as when the students \textit{adopted} a hint by first justifying it and then using it directly on their path to derive the conclusion. These two hint-specific metrics are the hint \textbf{Justification Rate} and \textbf{Adoption Rate}. 

\begin{figure*}[h!]
\centering
\includegraphics[width=\textwidth]{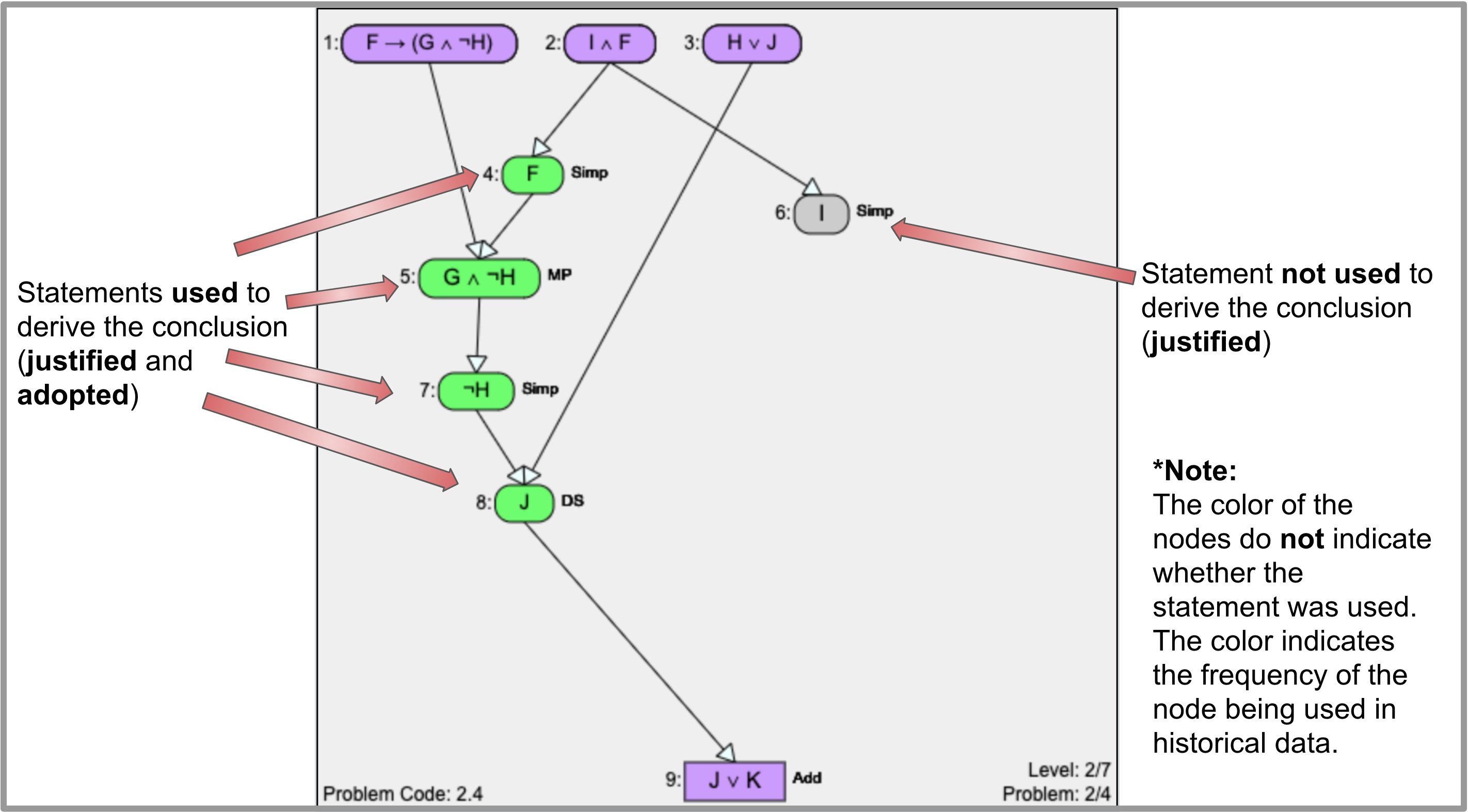}
\caption{A completed problem with nodes that were used to derive the conclusion (justified and adopted) and one node that was not used to derive the conclusion (justified but \textit{not adopted}). }
\label{fig:JustificationAdoption}
\end{figure*}

The hint Justification Rate is the percentage of unsolicited hints justified (correctly identifying the rule and prior nodes needed to derive the suggested node) divided by the total number of hints given across the training problems. A hint is said to be \textit{justified} when a student applies logic rules to existing logic statements to derive the hinted logic statement, and when a hint is justified, the tutor removes its `?' and connects it to its predecessor nodes with arrows labeled with the rule used to derive it. A hint justification provides evidence that a student noticed the hint and knew how to apply rules to justify it, but do not tell the full story. As in any problem-solving context, statements can be derived that are not needed in a final solution. Therefore, we also measure hint Adoption Rate, whether a hint contributes towards deriving the conclusion. A justified hint can be reached on a path from the problem's given statements. When a hint is \textit{adopted}, it must first be justified and then become necessary to a student's final solution -- in other words, the problem would be incomplete if the hinted statement were removed. This is shown visually when a directed path can be found from the hinted statement node to the problem conclusion. Figure \ref{fig:JustificationAdoption} shows a completed problem with labels indicating which nodes are considered \textit{justified} and which nodes were also \textit{adopted} for the solution. 

We also investigated impacts on help-seeking through the number of on-demand hint requests (when students click the``Get Suggestion” button). \textbf{Total Requests} represents the number of hint requests during the training portion. Data were analyzed to compare groups for the pretest, training, and posttest portions of the tutor. Within each hint group, we also compared performance of students with High or Low pretest scores, based on a median split on the pretest score.

To determine significant differences between hint types, we applied one way ANCOVA using the pretest as a covariate with Benjamini-Hochberg corrections to account for multiple tests. To check that the data met assumptions for ANCOVA, we used the the Shapiro-Wilk’s W test and Levene’s test, as well as visually inspecting the data via Q-Q plots and histograms. Data that did not meet the assumptions were transformed using log or square-root transformations, then re-inspected. Data reported in tables For clarity, all data in tables are reported before transformation.

\section{Results}

Table \ref{tab:suggestions} shows the overall hint metrics for each group during training. We expected \textit{Total Added}($F(2, 111)=160.20, p < 0.01$) and \textit{Steps Until Justified}($F(2, 111)=281.20, p < 0.01$) metrics to be significantly different, since each step of a problem can have a unique Next-Step (NS) but one Waypoint (WP) requires multiple steps to be derived. Based on prior literature on help avoidance and low help usage within tutors \cite{price2017factors}, we were pleasantly surprised to find students in both groups had relatively high justification and adoption rates. The Next-Step group justified a significantly ($F(2, 111)=12.96, p < 0.01$) higher percentage of hints, as shown by the \textit{Justification Rate}. Additionally, of the justified hints, we also saw a significantly ($F(2, 111)=5.49, p = 0.01$) lower Adoption Rate of the WP hints in students' final proofs, no significant interaction was observed with pretest proficiency. Although this is a relatively high number for both groups, the WP group's lower justification and adoption rates are concerning. 

This provides evidence in support of $H_{3}$ that Waypoint hints would be harder to derive; however, this evidence does not address whether this was due to the difficulty of the WP hints or students' lack of effort to derive them. We explore the possible reasons for these differences later in this section.

\begin{table}[H]
\caption{Hint metrics during training. For ANCOVA results controlling for the pretest score, p-values that are at least marginally significant are \textbf{bolded} and significant values also have an asterisk*.}
\label{tab:suggestions}
\centering
\begin{tabular}{llccc}
\hline
\multicolumn{2}{l}{}                                                     & \textbf{NS}             & \textbf{WP}             &            \\ \hline
\multicolumn{2}{l}{}                                                     & \textit{\textbf{n = 56}} & \textit{\textbf{n = 58}} &            \\ \hline
\multicolumn{2}{l}{\textit{\textbf{Metric}}}                             & \textit{Mean(SD)}        & \textit{Mean(SD)}        & \textit{p} \\ \hline
\rowcolor[HTML]{EFEFEF} 
\multicolumn{2}{l}{\cellcolor[HTML]{EFEFEF}Justification Rate} & 89\%(7)                  & 84\%(12)                 & \textbf{\textless{}0.01*}      \\
\multicolumn{2}{l}{Adoption Rate}                              &        83\%(10)                  &     74\%(17)                     & \textbf{0.01*}     \\
\rowcolor[HTML]{EFEFEF} 
\multicolumn{2}{l}{\cellcolor[HTML]{EFEFEF}Steps Until Justified}    & 1.1(0.1)                 & 2.2(0.3)                 & \textbf{\textless{}0.01*}   \\
\multicolumn{2}{l}{Total Added}                               & 49(9)                    & 30(7)                    & \textbf{\textless{}0.01*}  
\end{tabular}
\end{table}

To understand the overall impact of Next-Step versus Waypoint hints, we examined 
 the performance for both groups for the tutor pretest, training, and posttest, for all students regardless of incoming proficiency as shown in Table \ref{tab:pretrainingpost}. There were no significant differences between the WP and NS groups on the pretest, although the pretest was slightly worse for the NS group. During training, the NS group significantly outperformed the WP group with fewer steps, less time, better accuracy and overall score (Total Time $F(2, 111)=17.33, p\textless{}0.01$, Total Steps $F(2, 111)=3.73, p=0.02$, Accuracy $F(2, 111)=3.93, p=0.01$). The WP group, on average, took 20 minutes longer, took 36 more steps, and had 5\% lower accuracy on the training problems. No interactions were found between pretest proficiency and performance or hint metrics for the training.  
 
 These results suggest that Next-Step hints had a stronger impact than hypothesized in $H_{1}$ during training, with all Next-Step students outperforming all WP students. We examined help-seeking behaviors during training and found the NS group requested significantly more hints, although still a small number overall (approximately 1 per problem for NS versus 0.5 per problem on average for WP), so it is not likely that hint requests account for the difference in training performance. 

\begin{table*}[!h]
\centering
\caption{Performance metrics for each group on the pretest, training, and posttest; p-values that are at least marginally significant when applying ANCOVA controlled for pretest are \textbf{bold} and those that are significant also have an asterisk.}
\label{tab:pretrainingpost}
\begin{tabular}{llccc}
\hline
                  &                           & \textbf{NS (n = 56)}                         & \textbf{WP (n = 58)}                         & \multicolumn{1}{l}{}                         \\ \hline
                  & \textbf{Metric}           & \textit{Mean(SD)}                            & \textit{Mean(SD)}                            & \multicolumn{1}{l}{\textit{p-value}}         \\ \hline
\rowcolor[HTML]{EFEFEF} 
\textbf{Pretest}  & \textit{Total Time}       & 6.5(12)                                      & 5.5(9)                                       & 0.54                                         \\
                  & \textit{Total Steps}   & 19(35)                                       & 15(16)                                       & 0.84                                         \\
\rowcolor[HTML]{EFEFEF} 
                  & \textit{Accuracy}         & 67\%(22)                                     & 70\%(23)                                     & 0.57                                         \\
                  & \textit{}                 &                                              &                                              &                                              \\ \hline
\rowcolor[HTML]{EFEFEF} 
\textbf{Training} & \textit{Total Time (min)} & 58(24)                                       & 77(42)                                       & \textbf{\textless{}0.01*}                    \\
                  & \textit{Total Steps}   & 186(65)                                      & 222(99)                                      & \textbf{0.02*}                               \\
\rowcolor[HTML]{EFEFEF} 
                  & \textit{Accuracy}         & 74\%(10)                                     & 69\%(10)                                     & \textbf{0.01*}                               \\
                  & \textit{Total Requests}   & 15(29)                                       & 7(9)                                         & \textbf{0.04*}                               \\
\rowcolor[HTML]{EFEFEF} 
                  &                           & \multicolumn{1}{l}{\cellcolor[HTML]{EFEFEF}} & \multicolumn{1}{l}{\cellcolor[HTML]{EFEFEF}} & \multicolumn{1}{l}{\cellcolor[HTML]{EFEFEF}} \\ \hline
\rowcolor[HTML]{EFEFEF} 
\textbf{Posttest} & \textit{Total Time (min)} & 33(21)                                       & 42(34)                                       & \textbf{0.06}                                \\
                  & \textit{Total Steps}      & 99(40)                                       & 127(87)                                      & \textbf{0.02*}                               \\
\rowcolor[HTML]{EFEFEF} 
                  & \textit{Accuracy}         & 71\%(12)                                     & 66\%(10)                                     & \textbf{0.05*}                              
\end{tabular}
\end{table*}

More importantly, on the posttest, the NS group significantly outperformed the Waypoint group on Total Steps ($F(2, 111)=3.73, p=0.02$) and Accuracy ($F(2, 111)=2.38, p=0.05$). The WP group had 28 more total steps and had a 5\% lower accuracy, on average, on the posttest. There was a marginally significant difference between groups for the total time on the posttest ($F(2, 111)=2.78, p=0.06$), with the WP group spending roughly 10 more minutes on the posttest. No interactions were found between pretest proficiency and performance or hint metrics for the posttest. 

These results suggest that Next-Step hints had a stronger impact than hypothesized,  showing that, overall, the NS group performed better during training and the posttest. We believe that Next-Steps allow students to focus on solving one step, which we hypothesized would reduce time spent (since students did not have to determine \textit{what} to derive next when receiving hints, just the \textit{how}), and total steps (since the suggested hints were efficient).

\subsection{Effects on High- and Low- Pretest Groups}
Our hypotheses focused on the differential impact of hints based on incoming proficiency and the difficulty of applying Next-Step versus Waypoint hints. To investigate these hypotheses, we checked for differences in performance between prior proficiency groups within each group. We performed a median-split for incoming proficiency based on pretest scores and compared performance metrics across groups and proficiency (NS-High $n = 27$, WP-High  $n = 30$, NS-Low  $n = 29$, WP-Low  $n = 30$). 

First, we examined performance metrics for the High group, shown in Table \ref{tab:posttesthigh}. There were no significant differences between the NS and WP High groups on the pretest. For the training, the WP group took longer and made more mistakes, as indicated by the Total Time ($F(2, 54)=17.22, p < 0.01$), Accuracy ($F(2, 54)=5.291, p < 0.01$), and attempted more steps ($F(2, 54)=3.917, p = 0.03$). The hint justification rate was also significantly lower for the WP group ($F(2, 54)=4.49, p = 0.02$). No interactions were found between pretest proficiency and performance or hint metrics for the training. 

These results indicate that the WP-High group struggled with following hints, which may have led to them spending more time trying to figure out how to solve the problem. For the posttest, there were no significant differences between the NS-High and WP-High groups, although the WP-High group performed worse on average. This result confirms an aptitude-treatment interaction effect for high proficiency students where the treatment did not result in different results; i.e. high proficiency students were not as sensitive to the treatment choice (Next-Step or Waypoint). This means that $H_{2}$ was rejected; Waypoint hints did not improve performance for higher proficiency students.

\begin{table*}[]
\centering
\caption{Performance metrics between the NS and WP \textbf{High} proficiency groups for the pretest, training, and posttest of the tutor. ANOVA results are reported for the pretest. ANCOVA results, controlling for the pretest, are reported for the training and posttest, with p-values that are at least marginally significant in \textbf{bold} and significant p-values also have an asterisk *.}
\label{tab:posttesthigh}
\begin{tabular}{llccc}
\hline
\multicolumn{1}{l|}{\textbf{High Proficiency}} &                               & \textit{\textbf{NS-High}}                         & \textit{\textbf{WP-High}}                         & \multicolumn{1}{l}{}                          \\ \cline{1-1}
                                   & \multicolumn{1}{c}{\textit{}} & \textit{n = 27}                              & \textit{n = 30}                              & \textit{}                                     \\ \hline
                                   & \textit{\textbf{Metric}}      & \textit{Mean(SD)}                            & \textit{Mean(SD)}                            & \textit{p-value}                              \\ \hline
\rowcolor[HTML]{EFEFEF} 
\textbf{Pretest}                   & \textit{Total Time (min)}     & 1.7(0.64)                                    & 1.6(0.86)                                    & 0.12                                          \\
                                   & \textit{Total Steps}          & 5.6(1.7)                                     & 5.5(1.9)                                     & 0.68                                          \\
\rowcolor[HTML]{EFEFEF} 
                                   & \textit{Accuracy}             & 87\%(12)                                     & 87\%(14)                                     & 0.91                                          \\
                                   & \textit{}                     &                                              &                                              &                                               \\ \hline
\rowcolor[HTML]{EFEFEF} 
\textbf{Training}                  & \textit{Total Time (min)}     & 50(23)                                       & 64(41)                                       & \textbf{\textless{}0.01*}                     \\
                                   & \textit{Total Steps}          & 164(57)                                      & 198(86)                                      & \textbf{0.03*}                                 \\
\rowcolor[HTML]{EFEFEF} 
                                   & \textit{Accuracy}             & 73\%(10)                                     & 78\%(8)                                      & \textbf{\textless{}0.01*}                                \\
                                   & \textit{Total Requests}       & 8(8)                                         & 4(5)                                         & 0.22                                          \\
\rowcolor[HTML]{EFEFEF} 
                                   & \textit{Justification Rate}   & 89\%(7)                                      & 81\%(15)                                       & \textbf{0.02*}                     \\
                                   & \textit{Adoption Rate}        & \multicolumn{1}{l}{83\%(9)}                    & \multicolumn{1}{l}{\%74(17)}                   & \multicolumn{1}{l}{\textbf{\textless{}0.01*}} \\
\rowcolor[HTML]{EFEFEF} 
                                   &                               & \multicolumn{1}{l}{\cellcolor[HTML]{EFEFEF}} & \multicolumn{1}{l}{\cellcolor[HTML]{EFEFEF}} & \multicolumn{1}{l}{\cellcolor[HTML]{EFEFEF}}  \\ \hline
\rowcolor[HTML]{EFEFEF} 
\textbf{Posttest}                  & \textit{Total Time (min)}     & 33(21)                                       & 40(34)                                       & 0.29                                          \\
                                   & \textit{Total Steps}          & 81(48)                                       & 98(42)                                       & 0.18                                          \\
\rowcolor[HTML]{EFEFEF} 
                                   & \textit{Accuracy}             & 71\%(12)                                     & 67\%(9)                                      & 0.14                                         
\end{tabular}
\end{table*}

Next, we examined performance metrics for the Low pretest group. There were no significant differences between the NS and WP Low groups on the pretest. For the training, the WP took longer and attempted more steps, as indicated by the Total Time ($F(2, 54)=3.926 , p = 0.02$) and Total Steps ($F(2, 54)=12.96 , p < 0.01$). The hint justification rate was also significantly  lower for the WP group ($F(2, 54)=7.42, p < 0.01$). These results follow a similar pattern as the High group, in that the WP group performed worse overall in the training and were less able to justify the hints. For the posttest, the WP group continued the pattern of taking longer ($F(2, 54)=1.99, p = 0.09$) and attempting more steps ($F(2, 54)=3.93, p = 0.02$) with marginally significant and significant results, respectively, indicating that the (hypothesized) worse performance in the training portion may have transferred to their overall proof solving strategies on the posttest. No interactions were found between pretest proficiency and performance or hint metrics on the training or posttest.  

These results confirm hypothesis $H_{3}$ that Waypoints are more difficult for students and have a negative impact on training performance.

\begin{table*}[]
\centering
\caption{Performance metrics between the NS and WP \textbf{Low} proficiency groups for the pretest, training, and posttest. ANOVA results are reported for the pretest. ANCOVA results, controlling for the pretest are reported for the training and posttest; p-values that are at least marginally significant are in \textbf{bold} and significant p-values also have an asterisk *.}
\label{tab:posttestlow}
\begin{tabular}{llccc}
\hline
\multicolumn{1}{l|}{\textbf{Low Proficiency}} &                               & \textit{\textbf{NS-Low}}                         & \textit{\textbf{WP-Low}}                        & \textit{\textbf{}}                           \\ \cline{1-1}
                                  & \multicolumn{1}{c}{\textit{}} & \textit{n = 29}                              & \textit{n = 28}                              & \textit{}                                    \\ \hline
                                  & \textit{\textbf{Metric}}      & \textit{Mean(SD)}                            & \textit{Mean(SD)}                            & \textit{p-value}                             \\ \hline
\rowcolor[HTML]{EFEFEF} 
\textbf{Pretest}                  & \textit{Total Time (min)}     & 10(11)                                       & 11(15)                                       & 0.80                                         \\
                                  & \textit{Total Steps}          & 31(48)                                       & 25(19)                                       & 0.80                                         \\
\rowcolor[HTML]{EFEFEF} 
                                  & \textit{Accuracy}             & 49\%(13)                                     & 51\%(16)                                     & 0.88                                         \\
                                  &                               & \multicolumn{1}{l}{}                         & \multicolumn{1}{l}{}                         & \multicolumn{1}{l}{}                         \\ \hline
\rowcolor[HTML]{EFEFEF} 
\textbf{Training}                 & \textit{Total Time (min)}     & \textbf{66(23)}                              & \textbf{89(39)}                              & \textbf{0.02*}                               \\
                                  & \textit{Total Steps}          & \textbf{206(67)}                             & \textbf{249(107)}                            & \textbf{\textless{}0.01*}                               \\
\rowcolor[HTML]{EFEFEF} 
                                  & \textit{Accuracy}             & 69\%(10)                                     & 65\%(9)                                      & 0.13                                         \\
                                  & \textit{Total Requests}       & 13(13)                                       & 10(10)                                       & 0.55                                         \\
\rowcolor[HTML]{EFEFEF} 
                                  & \textit{Justification Rate}   & \textbf{90\%(8)}                             & \textbf{81\%(14)}                            & \textbf{\textless{}0.01*}                    \\
                                  & \textit{Adoption Rate}        & \textbf{83(10)}                              & \textbf{74(17)}                              & \textbf{0.02*}                               \\
\rowcolor[HTML]{EFEFEF} 
                                  &                               & \multicolumn{1}{l}{\cellcolor[HTML]{EFEFEF}} & \multicolumn{1}{l}{\cellcolor[HTML]{EFEFEF}} & \multicolumn{1}{l}{\cellcolor[HTML]{EFEFEF}} \\ \hline
\rowcolor[HTML]{EFEFEF} 
\textbf{Posttest}                 & \textit{Total Time (min)}     & \textbf{33(20)}                              & \textbf{43(35)}                              & \textbf{0.09}                                \\
                                  & \textit{Total Steps}          & \textbf{99(39)}                              & \textbf{128(94)}                             & \textbf{0.02*}                                \\
\rowcolor[HTML]{EFEFEF} 
                                  & \textit{Accuracy}             & 69\%(12)                                     & 65\%(10)                                     & 0.17                                        
\end{tabular}
\end{table*}

We hypothesized in $H_{1}$ that Next-Step hints would improve (training and posttest) performance compared to Waypoint hints, for low proficiency students. The overall performance (Table \ref{tab:pretrainingpost}) confirmed that the Next-Step hint group produced better training and posttest performance. However, Table \ref{tab:posttestlow} confirms that the benefits in the posttest are more prominently seen with the students with lower incoming proficiencies, confirming $H_{1}$.

 We hypothesized in $H_{3}$ that the Waypoint hints would cause lower justification rates and worse training performance due to their increased difficulty, which is seen with both the WP-High and WP-Low groups, confirming our $H_{3}$ hypothesis. There was also a significant difference in the Adoption rates between the NS and WP groups for both High and Low students, with the WP adoption rates being lower. This suggests that students were not, in fact, able to independently discover the strategies that underlie the WP hints.
 
 Although we expected a lower hint justification rate in the WP group, we thought that the increase in difficulty would be beneficial to high proficiency students by allowing them more exploration of the problem space. Therefore, we hypothesized in $H_{2}$ that higher incoming proficiency students would do better on the posttest after experiencing the WP hints in training. However, that is not the case. The WP-High group was only able to perform similarly to the NS-High group and overall performed worse, although not significantly. Therefore, $H_{2}$ is rejected. However, the results do seem to indicate that the high incoming-proficiency students were less affected by the treatment than the low incoming-proficiency students based on there being more significant results between conditions in the low incoming-proficiency group. As mentioned earlier, we expected that an aptitude-treatment interaction (ATI) might occur, where certain students are more sensitive to variations in the learning environment and may be affected differently by the treatment compared to less sensitive (more proficient) students who are able to perform well regardless of treatment.

\subsection{Did Waypoints help with strategy for those who could utilize them?}
Although the performance results caused us to reject $H_{2}$, we wanted to investigate whether WP hints provided strategy-related benefits to those students who were \textit{able} to use them. Therefore, we performed correlation analyses using the Pearson correlation coefficient between the hint Justification and Adoption rates with posttest performance metrics. For the correlation analyses, we used an R function, corr.test, which computes the Pearson correlation coefficient, significance tests using t-tests, and performs optional corrections which we specified as Bonferroni corrections\cite{Rpsch}.  Table \ref{tab:Corr} shows the significant correlations of hint Adoption and Justification Rates with performance metrics for NS and WP groups on the posttest, as well as correlations with the incoming proficiency groups. 

For the NS group, the only  significant correlation found was for the NS-High group between hint Adoption Rate and Total Steps ($p = 0.06$), showing a moderate, negative correlation. This could indicate that students in the NS-High group attempted fewer steps in the posttest (a better result) if they adopted more of the NS hints during training. For the WP group, there are moderate, negative correlations of Justification rate with Total Time ($p = 0.03$) and with Total Steps ($p = 0.01$), and also of Adoption Rate with Total Time ($p < 0.01$) and Total Steps ($p < 0.01$). So \textbf{justifying and adopting WPs were both associated with more efficient proofs that were shorter and achieved in less time}. There was also a significant moderate, negative correlation for the WP-Low group between Total Steps ($p = 0.04$) and the hint Adoption rate. For the WP-High group, there was a similar moderate, negative correlation  between Total Steps ($p = 0.02$) and hint Adoption Rate, but the WP-High group also had moderate, negative correlation between Total Time ($p = 0.02$) and  hint Adoption Rate.

This result aligns with our reasoning behind  $H_{2}$, that Waypoint hints should improve efficiency- and time-related metrics on the posttest, especially for higher proficiency students. However, ultimately, the WP students performed worse. Based on these results, we conclude that more support may be needed for WPs so that students can utilize them as well as NS hints to better achieve efficiency-related benefits. 

\begin{table}[h]
\centering
\caption{Significant correlations between hint Justification and Adoption rates with posttest performance metrics for each hint type group and pretest group}
\label{tab:Corr}
\begin{tabular}{lllll}
\hline
\textbf{Condition} & \textbf{Split} & \textit{\textbf{Metric-Pair}} & \multicolumn{1}{c}{\textit{Corr}} & \multicolumn{1}{c}{\textit{p}} \\ \hline
\rowcolor[HTML]{EFEFEF} 
\textbf{NS}       & High           & Adoption-Total Steps          & -0.38                             & 0.06                           \\ \hline
\textbf{WP}       & All            & Justification-Total Time (min)      & -0.30                             & \textbf{0.03*}      \\
\rowcolor[HTML]{EFEFEF} 
                   &                & Justification-Total Steps     & -0.32                             & \textbf{0.01*}      \\
                   &                & Adoption-Total Time (min)          & -0.35                             & \textbf{\textless{}0.01*}                 \\
\rowcolor[HTML]{EFEFEF} 
                   &                & Adoption - Total Steps        & -0.40                             & \textbf{\textless{}0.01*}                 \\ \cline{2-5} 
                   & High           & Adoption-Total Time (min)          & -0.40                             & \textbf{0.02*}                 \\
\rowcolor[HTML]{EFEFEF} 
                   &                & Adoption-Total Steps          & -0.41                             & \textbf{0.02*}                 \\ \cline{2-5} 
                   & Low            & Adoption-Total Steps          & -0.39                             & \textbf{0.04*}                
\end{tabular}
\end{table}

\subsection{What are the circumstances when hints were not used?}
To understand if the WP hints were actually harder to derive, as hypothesized $H_{3}$, we investigated how many unused (unjustified) hints were \textit{attempted} to be justified. The significantly lower difference in hint Justification Rate of the WP group as shown in Table \ref{tab:suggestions} and the significantly worse performance by the WP group in the training as shown in Table \ref{tab:pretrainingpost} led us to want to better understand the circumstances surrounding why the WP hints were used proportionately less. 

The hint Justification and Adoption rates can only tell us that students were, or were not, using the hints, but do not provide any insight into whether the students were actively attempting to derive the hint. Therefore, we conducted analyses to see if the WP hints were truly harder to derive ($H_3$). This would be indicated by the students attempting to work towards the hint, and not succeeding, versus outright ignoring the hint. Because the WP hints are more steps away than the NS hint, students see the hint as too complicated to be helpful and just ignore the hint outright. However, if students are attempting to to derive the hint and not able to be successful, this is a larger concern. To determine if students were attempting to derive the hint, we examined the steps taken after a hint was added (3 steps ahead for NS and 5 steps ahead for WP). If a majority of the steps examined contained variables that were also seen in the hint (2 out of 3 steps for NS and 3 out of 5 steps for WP) , it was considered \textbf{attempted}. 

Table \ref{tab:attempted} shows the total unused hints. \textbf{Total Unused} represents the total number of unused hints per person in each group. The \textbf{\% Unused/Total} is the total number of unused hints divided by the total number of hints that were added, which provides a clearer picture of the relative percentage of hints that were left unused by each student compared to how many they were being given. The \textbf{\% Attempted/Unused} is the total attempted hints divided by the total unused hints representing the percentage of the unused hints that were attempted. There was not a significant difference in the total amount of unused hints between the groups ($F(2, 111)=0.47, p = 0.49$). The WP group left a significantly larger percentage of  hints unused (\% Unused/Total) compared to the NS group ($F(2, 111)=12.96, p < 0.01$). Interestingly, the WP group were also attempting a larger amount of unused hints (\% Attempted/Unused) compared to the NS group ($F(2, 111)=3.013, p = 0.05$).

\begin{table*}[hb]
\centering
\caption{The total unused (unjustified) hints, percentage of hints unused out of all hints added, and the percentage of the unused hints that were attempted to be derived between the NS and WP group. For ANCOVA results controlling for the pretest score, p-values that are at least marginally significant are in \textbf{bold} and significant values also have an asterisk *.}
\label{tab:attempted}
\setlength{\tabcolsep}{8pt}
\begin{tabular}{lccc}
\hline
                             & \textbf{NS}              & \textbf{WP}              & \textbf{}                 \\ \hline
\rowcolor[HTML]{EFEFEF} 
\textit{\textbf{}}           & \textit{\textbf{n = 56}} & \textit{\textbf{n = 58}} & \textit{}                 \\ \hline
\textit{\textbf{Metric}}     & \textit{Mean(SD)}        & \textit{Mean(SD}         & \textit{p-value}          \\ \hline
\rowcolor[HTML]{EFEFEF} 
\textit{Total Unused}        & 5.4(4.4)                 & 6.6 (6.4)                & 0.49                      \\
\textit{\% Unused/Total}     & 10.4\%(7.2)              & 19.4\%(14.8)             & \textbf{\textless{}0.01*} \\
\rowcolor[HTML]{EFEFEF} 
\textit{\% Attempted/Unused} & 57.0\%(37.4)             & 72.6\%(28.9)             & \textbf{0.05*}            \\ \hline
\end{tabular}
\end{table*}

 To understand when unsolicited hints were \textbf{not} justified, we determined the circumstances when this occurred and illustrate several situations: when students attempted to use the hint, and what the eventual outcome was: either Gave Up or Solved Without using the hint. \textbf{Gave Up} represents any actions that end the problem without solving it, such as restarting or skipping the problem. In this situation, students had a hint on the screen, worked a few steps, then clicked the restart or skip button without justifying the hint. When a student clicks restart or skip, this erases all current progress on the problem. We considered this to be ``giving up" because the student is removing all progress made on the current problem by taking these actions, which is concerning given that a hint was on the screen. \textbf{ Solved Without} represents when students completed a proof with an unjustified hint still on the screen. In this case, students have a hint but eventually solve the problem without using the hint. This indicates that the hint was ignored, or at the very least, was not essential to solving the proof. We are less concerned with this case because the students were able to progress. However, since the hint is the most efficient step to work towards, any student who avoided it took a less efficient route to solve the problem. Lastly, although students had the option to delete a hint, no deletions were observed possibly due to students not knowing how to delete the hint. 

Table \ref{tab:twocases} details the two cases in which a hint was added, but the student did not justify it. For significant differences, ANCOVA was used with the pretest score as the covariate. The Total Unused, \% Unused/Total and the \% Attempted/Unused are defined above. We also examined how many steps the students took after a hint was given but before they gave up or solved the proof to determine how much effort was put into trying to derive the hint. \textbf{Steps Before} is the number of steps the student attempted \textit{after} receiving the hint \textit{before} they gave up or solved the proof. This metric was added to see how long students were trying to work on the problem after the hint was given. 
\begin{table}[h]
\centering
\caption{Comparison of unused hints of each subtype by amount, percentage that were attempted, and steps before the action occured.}
\label{tab:twocases}
\begin{tabular}{lcccc}
\hline
                                                               & \multicolumn{1}{l}{\textbf{Gave Up}} & \multicolumn{1}{l}{\textbf{}}                                      & \multicolumn{1}{l}{\textbf{Solved Without}} & \multicolumn{1}{l}{\textbf{}} \\ \hline
\rowcolor[HTML]{EFEFEF} 
\multicolumn{1}{c}{\cellcolor[HTML]{EFEFEF}\textit{\textbf{}}} & NS (n = 56)                          & \multicolumn{1}{c|}{\cellcolor[HTML]{EFEFEF}WP (n = 58)}           & NS (n = 56)                                 & WP (n = 58)                   \\ \hline
                                                               & \textit{Mean(SD)}                    & \multicolumn{1}{c|}{\textit{Mean(SD)}}                             & \textit{Mean(SD)}                           & \textit{Mean(SD)}             \\ \hline
\rowcolor[HTML]{EFEFEF} 
\textit{Total Unused}                                          & 2.9(3.5)                             & \multicolumn{1}{c|}{\cellcolor[HTML]{EFEFEF}4.0(5.20)}             & 2.5(1.8)                                    & 2.4(2.0)                      \\
\textit{\% Unused/Total}                                       & 46\%(34)                             & \multicolumn{1}{c|}{47\%(34)}                                      & 54\%(34)                                    & 53\%(34)                      \\
\rowcolor[HTML]{EFEFEF} 
\textit{\% Att./Unused}                                   & \textbf{48\%(43)}                & \multicolumn{1}{c|}{\cellcolor[HTML]{EFEFEF}\textbf{72\%(29)}} & 62\%(39)                                & 73\%(31)                  \\
\textit{Steps Before}                                          & 1.6(1.7)                             & \multicolumn{1}{c|}{2.4(2.0)}                                      & 4.6(2.4)                                    & 4.4(3.4)                     
\end{tabular}
\end{table}

The WP group attempted to derive a significantly higher number of the unused hints before giving up (\% Att./Unused: $F(2, 111)=2.75, p = 0.02$). There were no significant differences in the Total Unused for either cases (Gave Up: $F(2, 111)=1.86, p = 0.14$ and Solved Without: $F(2, 111)=0.67, p = 0.92$), the \% Attempted/Total for the Solved Without case ($F(2, 111)=1.718, p = 0.13$), or the Steps Before for either cases (Gave Up: $F(2, 111)= 1.141,p = 0.13$ and Solved Without: $F(2, 111)=0.172, p = 0.79$). Therefore, both groups of students had a similar distribution of unused hints in both cases; however, the WP group attempted to derive a significantly higher percentage of the hints in the Gave Up case. This result indicates that the WP students had attempted to make progress towards the hints, were unable to justify them, and then gave up. This is more concerning than giving up on a problem in which they had not attempted to derive the hint, and indicates the WP hints may have been too hard to derive. 

The purpose of this analysis was to investigate $H_{3}$ and determine the circumstances surrounding why the WP group had a significantly lower hint Justification Rate than the NS group. The results provide evidence in support of $H_{3}$ that the Waypoint hints were harder for students to derive.

\section{Discussion}

This work aims to explore the extension of a Next-Step hint generator to easily create subgoal-inspired assistance. The Next-Step group saw overall the best performance for both the training and posttest, including the students with lower incoming proficiencies providing supporting evidence for $H_{1}$. Our results indicated that the Waypoint group performed overall worse in both training and posttest causing us reject $H_{2}$. Results also showed that the lower proficiency students, specifically, were less able to utilize this form of assistance; however, students who were able to utilize Waypoints did see benefits in terms of time and efficiency on the posttest. Furthermore, we explored the circumstances surrounding when hints were not utilized and found that students in the Waypoint group attempted a larger percentage of the hints before giving up, providing evidence in support of $H_{3}$ that Waypoints would be harder to derive. In this section, we discuss the trade-offs of the two hint types.

\subsection{Waypoint hints}
WPs were intended for students to learn strategies for solving proofs by breaking the problem into smaller subgoals and providing students with more independent problem solving experience than NS. However, the majority of WP students appeared to have struggled with WP hints instead, a trade-off of the assistance dilemma \cite{koedinger2007exploring}. The WP group performed worse overall in both training and posttest portions of the tutor (see Table \ref{tab:pretrainingpost}). Another interesting result, shown in Section 5.1, is that the WP Low-pretest group has a significantly lower Justification Rate and marginally significant decrease in posttest performance metrics. This aligns with literature showing that lower proficiency learners are less able to use abstract guidance \cite{kirschner2006minimal, sweller2008evolutionary}. Therefore, the WP hints might not provide enough guidance for students. Research has shown that complex assistance can hinder learning by taxing cognitive load \cite{sweller1988cognitive, sweller2011cognitive}, which can happen when learners try to process new information and incorporate complex assistance at the same time and ``thus forcing learners to use random search procedures" \cite{kalyuga2007enhancing}.  This is a limitation of our study as Waypoints may produce better results with more scaffolding.

The Justification Rate being significantly lower for the WP group indicates that the lower performance may be due to an inability to properly use the assistance (see Table \ref{tab:suggestions}). This is partially supported by Table \ref{tab:attempted} and Table \ref{tab:twocases}, which shows that the WP group had a higher percentage of attempts to justify a hint without succeeding, compared to the NS group.  The Adoption Rate being significantly lower for the WP group indicates that, even when students in the WP group were able to justify the hints, they were less able to adopt them to connect the WP hints to the conclusion. Due to the design of the hint being a few steps away, students could end up on a solution path different from the path initially given by the WP. Consequently, students who were unable to justify the WP or adopt it into the solution were not following the most efficient path, hindering their ability to learn from the strategies behind the WP hints. 

One potentially positive result with Waypoints is shown in Table \ref{tab:Corr}, with respect to the significant negative correlations of Justification and Adoption rate with total time and total steps on the posttest. Students who were able to justify and adopt the WPs were associated with taking a shorter time and fewer steps on the posttest. This correlation aligns with our original intention of using WP to support strategy development by helping students become more efficient in their problem solving process. Therefore, it is possible that students with more experience and domain knowledge may better utilize Waypoints and receive strategy-related benefits. However, it is important to take into account that correlation analyses cannot determine causality and there could be variables not included in these analyses that play an important role in these relationships \cite{fields2012discovering}. Therefore, this interpretation is only a possibility. Based on these results, WPs can be improved by providing more information (perhaps automatically provided once we detect that a student is unsuccessfully attempting to justify the hint) or incorporating ideas from recent research with promising methods of scaffolding goal-based hints \cite{marwan2019impact}. 

\subsection{Next-Step hints}
The total time, total steps, and accuracy were significantly different, or trending towards significance, between groups as shown in Table \ref{tab:pretrainingpost} for the training and posttest. Since the groups had similar pretest scores, these results show that both the NS and WP groups came into the tutor performing similarly, but by the posttest the two groups had diverged; the NS group had higher accuracy and fewer total steps. Furthermore, the NS group were able to increase their accuracy between the non-isomorphic pre- and posttest compared to the WP group who did not show such improvements. This was perhaps due to the increased practice in applying rules to justify both unsolicited and on-demand hints - since the NS group received and justified significantly more hints in both of these categories.

The differences in time, steps, and accuracy between the groups show that NSs were more beneficial for students. As shown in Table \ref{tab:suggestions}, there is a significant difference in the higher Justification Rate for NSs. We believe these results may be due to the alignment of NSs with novice's bottom-up problem solving approaches that focus on what to do in the short term \cite{anderson1984learning, guzdial1995centralized, sweller2008evolutionary}. NSs may also have potentially resulted in an overall lower cognitive load \cite{kalyuga2011cognitive}, though this supposition is only based on their design and not data from students. As a justification, students considering NSs only needed to think about which nodes and rules could be used to derive the NS. In contrast, WP students needed to think about which nodes to use, which rules to apply, and what intermediate steps they would have to achieve before deriving the WP. 

Interestingly, the NS group requested more on-demand hints (see \ref{tab:pretrainingpost}). This suggests that the NS group may have found the assistance more helpful and became more comfortable requesting help. Prior research has shown students are more likely to request help when they received help that they perceive to be more suitable for their needs \cite{price2017factors}. 

Although WPs were designed to promote more independent, strategic problem solving, it is possible that NSs also helped students learn problem solving strategies. Based on the hint generator design, NS students following the hints were seeing the most efficient next step based on the current proof state. Problems with frequent Next-Step hints could be acting as partially-worked examples, which are known to increase efficient problem solving strategies \cite{sweller1985use, mclaren2014exploring}. Previous research on hint usage during problem solving in programming suggests that hints can, sometimes, save students time but reduce learning \cite{morrison2015subgoals}. In our research, NS hints seem to save students time and increase performance on the post-test. This suggests that NS hints may help students learn to solve the problems more efficiently (more quickly and with fewer steps).

\section{Conclusion}
This paper contributes a study showing an extension of the Hint Factory to create higher-level hints, and the effects of two types of hints on students' efficiency and accuracy in solving logic proofs: Next-Step hints (NSs) and Waypoint (WPs) hints. It is important to note these hints were provided unsolicited as well as through on-demand hint requests, which could affect the students' usage and reception of the hints. Furthermore, our hints are provided periodically and not necessarily when a student may need them. However, our prior has shown our unsolicited, periodically provided hints do not have any negative impacts on training or post-test performance metrics compared to students in the only on-demand hint group. In this paper, NSs helped students become quicker, more accurate, and more efficient in their proofs. However, the more distant goals of WPs seemed to be harder for the students, which not only affected the training problems where the assistance occurred, but resulted in lower accuracy and reduced efficiency in the posttest. Despite the WP group spending more time on problem solving during training, their performance did not benefit as much as the NS group. Furthermore, learners with lower incoming proficiency were least able to utilize WPs, while NSs provided benefits to both higher and lower proficiency groups. Although NS performed better overall, students who were able to incorporate WP, especially those in the WP-High group, saw benefits in terms of time and efficiency on the posttest.

Another interesting outcome was that the NS group had higher justification rates and requested more help, which agrees with previous research showing that hint quality affects help-seeking behaviors. In the future, WPs could be augmented to reduce cognitive load without eliminating the multi-step aspect by eliminating other elements of the task, such as highlighting needed nodes or offering multiple hint levels. Other future work includes using machine learning to determine when to provide a hint rather than providing them periodically. Finally, we hope to transfer these findings to other open-ended problem domains like programming in order to offer additional instructional supports and hints to novice students.

\bibliographystyle{spmpsci}
\bibliography{Manuscript}

\end{document}